\documentstyle[12pt,axodraw]{article} 
\catcode`@=11
\newcount\@tempcntc
\def\@citex[#1]#2{\if@filesw\immediate\write\@auxout{\string\citation{#2}}\fi
\@tempcnta\z@\@tempcntb\m@ne\def\@citea{}\@cite{\@for\@citeb:=#2\do
{\@ifundefined
{b@\@citeb}{\@citeo\@tempcntb\m@ne\@citea\def\@citea{,}{\bf
?}\@warning
       {Citation `\@citeb' on page \thepage \space undefined}}%
    {\setbox\z@\hbox{\global\@tempcntc0\csname b@\@citeb\endcsname\relax}%
     \ifnum\@tempcntc=\z@ \@citeo\@tempcntb\m@ne
       \@citea\def\@citea{,}\hbox{\csname b@\@citeb\endcsname}%
     \else \advance\@tempcntb\@ne \ifnum\@tempcntb=\@tempcntc
      \else\advance\@tempcntb\m@ne\@citeo
      \@tempcnta\@tempcntc\@tempcntb\@tempcntc\fi\fi}}\@citeo}{#1}}
\def\@citeo{\ifnum\@tempcnta>\@tempcntb\else\@citea\def\@citea{,}%
  \ifnum\@tempcnta=\@tempcntb\the\@tempcnta\else
   {\advance\@tempcnta\@ne\ifnum\@tempcnta=\@tempcntb \else
   \def\@citea{--}\fi
   \advance\@tempcnta\m@ne\the\@tempcnta\@citea\the\@tempcntb}\fi\fi}
   \catcode`@=12

\begin{document}
\pagestyle{empty}
\begin{flushright}
   CERN-TH/98-343\\[-0.2cm]
   FERMILAB-PUB-98/347-T\\[-0.2cm]
   UICHEP-TH/98-12\\[-0.2cm]
   hep-ph/9811202\\[-0.2cm]
   October 1998
\end{flushright}

\begin{center}
{\Large {\bf New two-loop contribution to electric dipole moment in
supersymmetric theories}}\\[1.cm]
{\large Darwin Chang$^{a,b}$, Wai-Yee Keung$^{c,b}$, 
and Apostolos Pilaftsis$^{d,b}$}\\
{\em $^a$NCTS and Physics Department, National Tsing-Hua University,\\
Hsinchu 30043, Taiwan, R.O.C.}\\
{\em $^b$Fermilab, P.O. Box 500, Batavia IL 60510, U.S.A.}\\
{\em $^c$Physics Department, University of Illinois at Chicago, IL
  60607-7059,  USA}\\
{\em $^d$Theory Division, CERN, CH-1211 Geneva 23, Switzerland}
\end{center}
\bigskip\bigskip
\centerline{\bf  ABSTRACT} 
We  calculate  a new type of  two-loop  contributions to  the electric
dipole moments of the electron and neutron in supersymmetric theories.
The new contributions are originated from the  potential CP violation in
the trilinear couplings of the Higgs bosons to the scalar-top or the 
scalar-bottom quarks.  These  couplings  were previously very weakly
constrained.   The electric dipole moments are induced  through a
mechanism  analogous to that due to  Barr and Zee.  We find observable
effects for a sizeable portion of the parameter space related to the third
generation scalar-quarks in the minimal supersymmetric standard model
which cannot be excluded by earlier considerations.\\[0.3cm]
PACS numbers: 11.30.Er, 14.80.Er

\newpage
\pagestyle{plain}
Supersymmetry (SUSY) is considered to be  theoretically   the most
conceivable avenue   known so  far  which  may  lead to  a  successful
unification of  gravity with all other fundamental  forces by means of
supergravity and superstrings.  SUSY also  provides the most appealing
perturbative solution to  the gauge-hierarchy and naturalness problems
of the standard model (SM).  However, SUSY is not an exact symmetry of
nature.  Its minimal realization, the minimal supersymmetric SM (MSSM)
must break SUSY  softly  in    order  to accomplish agreement     with
experimental observations.  The  scale of SUSY-breaking  should not be
much higher than few TeV if one wishes to retain  the good property of
perturbative naturality mentioned above, namely quantum corrections to
the parameters of  the theory must not  exceed in size  the parameters
themselves up to energies of  the Planck  scale.   On the other  hand,
unlike the SM,  a serious weakness of  supersymmetric theories as well
as of the MSSM \cite{GD,NS} is their failure  to explain the smallness
of the observed flavour-changing neutral currents (FCNC) involving the
first  two families of  quarks,  and the  absence of sizeable electric
dipole moments (EDMs)   of the neutron   and electron \cite{BS}.   The
present   experimental upper  bounds  on  the neutron   EDM $d_n$  and
electron EDM $d_e$ are very tight  \cite{PDG}: $|d_n| < 10^{-25}\ e$cm
and $|d_e| < 10^{-26}\ e$cm.

As a result of the afore-mentioned FCNC and  CP crises, some degree of
fine-tuning  is necessary in generic  supersymmetric theories to avoid
these problems.  A  few  phenomenologically attractive solutions  have
been  suggested  in  the literature.    For example, Ref.\  \cite{CKN}
suggests a  solution within  the  context of  an effective  SUSY model
which seems  to  combine all healthy   features of both the   MSSM and
technicolour theories.  The main virtue of the effective SUSY model is
that any  non-SM source of  CP violation and  FCNC involving the first
two  generations    is   suppressed  by  allowing     their respective
soft-SUSY-breaking masses  to be as   high as  20  TeV, whereas  third
generation scalar quarks and leptons may naturally be light well below
the TeV scale.  Another interesting way to suppress CP-conserving FCNC
interactions without resorting to a  high scale of SUSY-breaking is to
impose  a  kind  of   universality \cite{GD}   or  alignment \cite{NS}
condition on   the flavour  space    of all   scalar fermions.    Most
interestingly, we shall see  that the existing bounds  on CP-violating
operators such  as EDMs of the  electron and  neutron are dramatically
relaxed if  an approximate alignment  between the  $\mu$ parameter and
the gaugino  masses exists and  the  trilinear couplings $A_f$  of the
Higgs-boson to the  first two generations are  very  small.  Note that
these  suppression mechanisms for EDMs  are  designed to suppress only
those contributions that do not involve the third generation.

Whichever the  suppression mechanism of FCNC and  CP violation for the
first two families might be, large CP-violating trilinear couplings of
the Higgs bosons  to the scalar-top  and scalar-bottom quarks can lead
by themselves to  large loop effects  of CP noninvariance in the Higgs
sector of the MSSM \cite{APLB}.  In this Letter we shall show that the
very same source of CP violation can give rise to EDMs of electron and
neutron at the observable level 
by virtue of the two-loop  Barr-Zee-type graphs \cite{BZ,CKY} shown in
Fig.\ 1.  Apart   from the MSSM,  these  novel  EDM contributions  are
present in any supersymmetric theory  and as we  will see, for a  wide
range  of parameters  they may  even  dominate  by several orders   of
magnitude over  all other  one-,   two- and  three-loop  contributions
discussed      extensively      in   the     existing       literature
\cite{EFN,KO,FOS,IN,TKNO,DDLPD,FPT}.

Supersymmetric theories contain many new CP-odd phases that are absent
in the SM.  However, not  all of them  are physical.  Specifically, in
the MSSM  \cite{EFN} supplemented by a  universality condition  at the
grand unification  scale, only two of  the four complex parameters $\{
\mu ,\ B,\ m_\lambda  ,\ A \}$ are phase  convention independent.  For
instance,  one can  absorb  the  common phase of   the  gaugino masses
$m_\lambda$  by a  chiral  rotation into the   $\lambda$ fields, where
$\lambda$  collectively    represents  the   gauginos     $\tilde{g}$,
$\tilde{W}$ and  $\tilde{B}$ of the  SU(3)$_c$, SU(2)$_L$ and U(1)$_Y$
gauge   groups,  respectively.    Furthermore,  the    renormalization
condition  that  the total tadpole  contribution  to the  CP-odd Higgs
boson  $a$ must  vanish  \cite{APLB} relates the    phase of the  soft
bilinear Higgs-mixing mass $B\mu$  to the phase  difference of the two
Higgs doublets   in the MSSM  order by   order in perturbation theory,
where  $\mu$ is  the  usual mixing parameter   of the two Higgs chiral
superfields in the  superpotential.   In  accordance with the    above
renormalization condition, one  may then choose  a basis for the Higgs
fields in  which $B\mu$ is real  at the tree level.  Consequently, the
two non-trivial CP-violating   phases in the  weak  basis  under study
reside in $\mu$ and the universal soft trilinear coupling $A_f = A$ of
the Higgs fields to the scalar fermions $\tilde{f}$.

Before we calculate the two-loop Barr-Zee-type diagrams in Fig.\ 1, we
shall briefly review the   most significant one-, two- and  three-loop
contributions to the  electron and neutron EDMs in  the MSSM.   At one
loop,  the dominant contributions  to the EDMs   of electron and $u$-,
$d$-    quarks   come     from    the  chargino     mass   eigenstates
$\tilde{\chi}^+_1$, $\tilde{\chi}^+_2$  and  the gluino   $\tilde{g}$,
respectively    \cite{EFN,KO,FOS}.        Chargino   quantum   effects
 give rise to a EDM of a fermion $f$
\begin{equation}
  \label{1chargino}
\Big( \frac{d_f}{e}\Big)^{\tilde{\chi}^+}\ \approx\  
\frac{\alpha_w}{2\pi}\, R_f\, T^f_z\
 \frac{ m_{{}_{\widetilde{W}}} {\rm Im}\ \mu }{M^2_{\tilde{\chi}^+_1} 
- M^2_{\tilde{\chi}^+_2}}\ 
\frac{m_f}{M^2_{\tilde{f}'} }\  \Big[
J\Big(\, \frac{m^2_{\tilde{\chi}^+_2}}{M^2_{\tilde{f}'}}\,\Big)\ -
J\Big(\, \frac{m^2_{\tilde{\chi}^+_1}}{M^2_{\tilde{f}'}}\,\Big)\, \Big]\, ,
\end{equation}
where  $\alpha_w=g^2_w/(4\pi)$     is  the  SU(2)$_L$   fine-structure
constant,  $R_f = \cot\beta\ (\tan\beta)   $ for $T^f_z= 1/2\ (-1/2)$,
$\tilde{f}'$ denotes a scalar fermion  having opposite weak isospin to
$\tilde{f}$, {\em i.e.}, $T^f_z =  - T^{f'}_z$, and  $J (z) = [3/2\ -\ 
z/2\ +\ \ln z/(1-z) ]/(1-z)^2$  is a one-loop function \cite{KO}, with
$J(1) = -1/3$, $J(z\ll 1) = (3 + 2\ln z)/2$ and $J(z\gg 1) = -1/(2z)$.
Correspondingly, the gluino contribution to the EDM  of a quark $q$ is
given by
\begin{equation}
  \label{1gluino}
\Big( \frac{d_q}{e}\Big)^{\tilde{g}}\ =\ \frac{2\alpha_s}{3\pi}\,
Q_q\, \frac{ {\rm Im}(A_q + R_q \mu^* )}{M_{\tilde{q}} }\ 
\frac{m_q}{M^2_{\tilde{q}} }\ \frac{m_{\tilde{g}} }{M_{\tilde{q}}}\ 
K \Big(\, \frac{m^2_{\tilde{g}}}{M^2_{\tilde{q}}}\,\Big)\, ,
\end{equation}
where  $\alpha_s=g^2_s/(4\pi)$ is  the strong fine-structure constant,
$Q_q$ is the charge  of (scalar) quarks  in $|e|$ units ($Q_u  = 2/3$,
$Q_d =   -1/3$),   and  $K(z)=    -  [1/2\   +\  5z/2\  +\   z(2+z)\ln
z/(1-z)]/(1-z)^3$  \cite{KO};  $K(1)  =  -1/12$; $K(z\ll  1)  = -1/2$;
$K(z\gg 1) = (5/2 - \ln z)/z^2$.  Recently, one-loop chromo-EDM (CEDM)
contributions  have  been studied   in   \cite{IN}  and  found to   be
comparable to  the EDM ones  with related  dependence.   A fairly good
estimate of the  combined effect of  all one-loop contributions to the
electron  and $u$-,   $d$-  quark EDMs, including  neutralino  quantum
corrections, may generically be obtained by
\begin{equation}
  \label{estim}
\Big( \frac{d_f}{e}\Big)^{\rm 1-loop}\ \sim\ 10^{-25}\ {\rm cm}\times
\frac{\{ {\rm Im}\ \mu ,\ {\rm Im}\  A_f \} }
{\max (M_{\tilde{f}}, m_{\lambda} )}\ 
\Big(\, \frac{1\ {\rm TeV}}{\max (M_{\tilde{f}}, m_{\lambda} )}\,\Big)^2
\Big(\, \frac{m_f}{10\ {\rm MeV}}\,\Big)\ .
\end{equation}
{}From Eq.\  (\ref{estim}) one finds the   known result \cite{KO} that
large  CP-violating phases are only  possible if scalar  quarks of the
first   two  families or  gauginos    have  few TeV masses.    Another
interesting  and, perhaps more  natural, way  to suppress the one-loop
EDM contributions having most of the SUSY particles much below the TeV
scale is to require  ${\rm  arg}(\mu) \stackrel{\displaystyle <}{\sim  }
10^{-2}$,  a constraint also  favoured  by cosmological considerations
\cite{FOS},  and assume  an  hierarchic  pattern for $A_f$'s:  $A_e\,,
A_{u,c}\,, A_{d,s} \stackrel{\displaystyle  <}{\sim } 10^{-3}\,  \mu$,
whereas $A_t$ and $A_b$ are the only  large trilinear couplings in the
theory with CP-violating phases of order unity.

At  two loops, neutron and  electron  EDMs receive a contribution from
the  $W$-boson EDM  through a one-loop   graph mediating charginos and
neutralinos \cite{TKNO}.  However,  these effects were  found to be at
least  one order  of magnitude  smaller  than the present experimental
bounds if  $\mu$,  $\tilde{m}_2 \sim  100$  GeV, and become more  than
two-orders suppressed if $\mu$, $\tilde{m}_2  > 300$ GeV.  The results
are  independent of the scalar  fermion masses but  depend linearly on
${\rm arg}\mu$, and therefore, the suppression is substantial if ${\rm
  arg}(\mu) \sim 10^{-2}$.  As  we will discuss  below, apart from the
two-loop small effects, there are  sizeable contributions to EDMs  due
to two-loop Barr-Zee-type  graphs.   Finally, there  is  a significant
three-loop contribution  to neutron EDM through Weinberg's three-gluon
operator \cite{SW,DDLPD}.  Recent  studies  have shown \cite{IN}  that
these   effects  scaling as  $1/m_{\tilde{g}}^3$  are   well below the
experimental neutron  EDM bound if gluinos are  heavier than about 400
GeV.

To  demonstrate explicitly  the significance of  the new Barr-Zee-type
contribution displayed in  Fig.\ 1, we  shall adopt a SUSY scenario in
which the only large CP-violating phase is contained  in $A_\tau = A_t
=  A_b$.   As has   been discussed above,    such  a scenario  is also
compatible with present experimental upper  bounds  on EDMs.  In  this
model, CP  violation is induced by   the interaction Lagrangian having
the generic form
\begin{equation}
  \label{Lcp}
{\cal L}_{\rm CP}\ =\ -\, \xi_f v\, a\, (\tilde{f}_1^* \tilde{f}_1\, -\,
\tilde{f}_2^* \tilde{f}_2)\
+\ \frac{ig_w m_f}{2M_W}\, R_f\, a\,\bar{f}\gamma_5 f\, ,
\end{equation}
where $a$ is  the would-be CP-odd Higgs  boson, $M_W = {1\over2}g_w v$
is the  $W$-boson mass, and $\tilde{f}_1$,  $\tilde{f}_2$ are  the two
mass-eigenstates    of the left-handed    and  the right-handed scalar
fermions  of the third  family.  Moreover,  $\xi_f$  is a CP-violating
parameter  depending on the  supersymmetric model under consideration. 
{}From    all     scalar   fermions,    only   $\tilde{t}_{1,2}$   and
$\tilde{b}_{1,2}$   are  Yukawa-coupling enhanced,   and are therefore
expected to give the biggest  contributions.  Another important  point
is  that   CP  violation induced  by    the  Lagrangian (\ref{Lcp}) is
proportional to the mass difference of scalar quarks, $m_{\tilde{f}_1}
- m_{\tilde{f}_2}$, and is  big  for the states $\tilde{t}_{1,2}$  and
$\tilde{b}_{1,2}$ as their respective mass  splitting may naturally be
maximal.  In the MSSM, the quantities $\xi_t$ and $\xi_b$ are given by
\cite{APLB}
\begin{equation}
  \label{xiq}
\xi_t\ =\ \frac{\sin 2\theta_t m_t {\rm Im} ( \mu
  e^{i\delta_t})}{\sin^2\beta\ v^2}\,,\qquad
\xi_b\ =\ \frac{\sin 2\theta_b m_b {\rm Im} ( A_b
  e^{i\delta_b})}{\sin\beta\, \cos\beta\, v^2}\, ,
\end{equation} 
where $\delta_q = {\rm arg} (A_q + R_q \mu^*)$. 
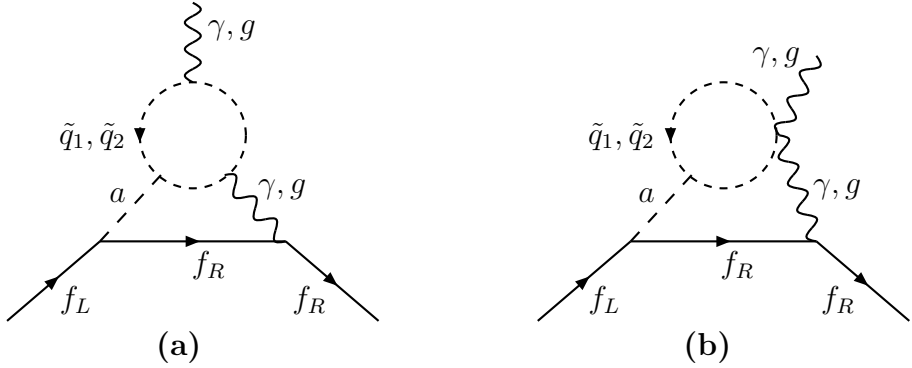
\begin{figure}

\begin{center}
\begin{picture}(400,200)(0,0)
\SetWidth{0.8}
 
\ArrowLine(10,30)(45,60)\Text(30,35)[lb]{$f_L$}
\DashLine(45,60)(68,85){5}\Text(55,75)[rb]{$a$}
\Photon(92,85)(115,60){3}{3}\Text(105,80)[l]{$\gamma,g$}
\ArrowLine(45,60)(115,60)\Text(80,51)[l]{$f_R$}
\ArrowLine(115,60)(150,30)\Text(118,35)[lb]{$f_R$}
\Photon(80,120)(80,150){3}{3}\Text(86,140)[l]{$\gamma,g$}
\DashArrowArc(80,100)(20,0,360){3}\Text(55,100)[r]{$\tilde{q}_1,\tilde{q}_2$} 

\Text(75,20)[]{\bf (a)}

\ArrowLine(210,30)(245,60)\Text(230,35)[lb]{$f_L$}
\DashLine(245,60)(268,85){5}\Text(255,75)[rb]{$a$}
\Photon(301,100)(315,60){3}{4}\Text(315,80)[l]{$\gamma,g$}
\Photon(301,100)(315,130){3}{3}\Text(292,130)[l]{$\gamma,g$}
\ArrowLine(245,60)(315,60)\Text(280,51)[l]{$f_R$}
\ArrowLine(315,60)(350,30)\Text(318,35)[lb]{$f_R$}
\DashArrowArc(280,100)(20,0,360){3}
\Text(255,100)[r]{$\tilde{q}_1,\tilde{q}_2$} 

\Text(275,20)[]{\bf (b)}

\end{picture}
\end{center}
\vspace{-1.cm}
\caption{Two-loop contribution to EDM and CEDM in supersymmetric
  theories (mirror graphs are not displayed.)}\label{f1}
\end{figure}

We   shall now   consider  the $\tilde{t}$-  and  $\tilde{b}$-mediated
two-loop graphs shown in Fig.\ 1  which may give rise to  an EDM and a
CEDM   for a light    fermion.  There  is in   principle  an analogous
contribution  to EDM due   to  a chargino loop.   Such  a contribution
originating  from   the     CP-violating   part    of    the  coupling
$a\chi^+_i\chi^-_i$ ($i=1,2$)  is proportional to  ${\rm arg}\mu$, and
hence small in our model.  We  also neglect Barr-Zee-type graphs where
the  photon is replaced by  a $Z$ boson, as  the vectorial part of the
$Z$-boson-mediated interaction is  suppressed relative to the photonic
one  by a factor $(1 -  4\sin^2\theta_w )/4 \approx 2.4\ 10^{-2}$ with
$\cos\theta_w = M_W/M_Z$ for the electron case,  and roughly $1/4$ for
the $u$ and  $d$ quarks.  Under  these  assumptions, a straightforward
calculation of the EDM  of a light  fermion $f$ induced by Figs.\ 1(a)
and 1(b) at the electroweak scale yields
\begin{equation} 
  \label{EDMf}   
\left( \frac{d_f}{e}\right)^{\tilde{q}}_{\rm EW} \ =\ Q_f\,   
\frac{3\alpha_{\rm em}}{64\pi^3}\, \frac{R_f\, m_f}{M^2_a}
 \sum_{q = t,b}\ 
\xi_q\, Q^2_q\,\left[\, 
F\left({M^2_{\tilde{q}_1} \over M^2_a}\right)
 -
F\left({M^2_{\tilde{q}_2} \over M^2_a}\right) \right]\, ,
\end{equation}
where  $\alpha_{\rm   em} = e^2/(4\pi)$   is  the electromagnetic fine
structure   constant, the subscripts EW    indicate that all kinematic
parameters   must  be evaluated at the    electroweak scale $M_Z$, and
$F(z)$ is a two-loop function given by
\begin{eqnarray} 
  \label{Fz}
F(z) &=& \int_0^{1} dx\ \frac{x(1-x)}{z - x(1-x)}\ 
\ln \Big[\,\frac{x(1-x)}{z}\,\Big]\nonumber\\
&& \hspace{-1.6cm}=\ 2 + \ln z + \frac{z}{x_+ - x_-}\, \Big\{ 
\Big[\, \ln\Big(\!-\frac{x_+}{x_-}\Big)\, -\,
\ln\Big(\!-\frac{x_-}{x_+}\Big)\, \Big]
\ln z\, +\, 2{\rm Li}_2 \Big(\frac{1}{x_+}\Big)\, -\, 
2{\rm Li}_2 \Big(\frac{1}{x_-}\Big)\,\Big\}\, ,\quad
\end{eqnarray}
with $x_\pm =  {1\over2}  (1 \pm  \sqrt{1-4z})$ and the  dilogarithmic
function defined as ${\rm   Li}_2  (z) =  \int_0^1   dt\, \ln t/[t   -
(1/z)]$; $F(z\ll  1) = \ln z  + 2$; $F(z\gg  1) =  - {1\over6}(\ln z +
{5\over3} )/z$.  In addition, the CEDM at the electroweak scale reads
\begin{equation} 
  \label{CEDMf}   
\left( \frac{d^C_f}{g_s}\right)^{\tilde{q}}_{\rm EW} =\
\frac{\alpha_s}{128\pi^3}\, \frac{R_f\, m_f}{M^2_a}\ \sum_{q = t,b}\ 
\xi_q\, \left[ 
F\left( {M^2_{\tilde{q}_1} \over M^2_a} \right) - 
F\left( {M^2_{\tilde{q}_2} \over M^2_a} \right) 
       \right]\, .
\end{equation}
The neutron EDM $d_n$ induced  by $d_u$ and  $d_d$ may be estimated in
the valence quark model at  the hadronic scale $\Lambda$ including QCD
renormalization effects \cite{CKY} through the expression
\begin{equation}
  \label{dne}
\frac{d_n}{e}\ =\ 
\left(\frac{g_s (M_Z )}{g_s (m_b)}\right)^{32\over23}
\left(\frac{g_s (m_b )}{g_s (m_c)}\right)^{32\over25}
\left(\frac{g_s (m_c )}{g_s (\Lambda)}\right)^{32\over27}
\left[\, \frac{4}{3}\,\left( \frac{d_d}{e}\right)_\Lambda
 - \frac{1}{3}\,\left( \frac{d_u}{e}\right)_\Lambda\, \right]\, .
\end{equation}
Here anomalous dimension factors  are  explicitly separated out   from
quantities $(d_d/e)_\Lambda$  and $(d_u/e)_\Lambda$  which  are simply
given by Eq.\  (\ref{EDMf}) with the running couplings and the running
masses of $u$- and $d$-quarks evaluated at the low-energy  hadronic scale
$\Lambda$.  For
definiteness, we take $m_u (\Lambda ) = 7$ MeV,  $m_d (\Lambda ) = 10$
MeV, $\alpha_s (M_Z) = 0.12$, and $g_s (\Lambda )/(4\pi) = 1/\sqrt{6}$
\cite{SW}. By analogy, the light-quark CEDMs  $d_u^C$ and $d^C_d$ give
rise to a neutron EDM
\begin{equation}
  \label{cdne}
\frac{d^C_n}{e}\ 
=
\left(\frac{g_s (M_Z )}{g_s (m_b)}\right)^{28\over23}
\left(\frac{g_s (m_b )}{g_s (m_c)}\right)^{28\over25}
\left(\frac{g_s (m_c )}{g_s (\Lambda)}\right)^{28\over27}
\left({g_s(M_Z)\over g_s(\Lambda)}\right)^2
\left[\,  \frac{4}{9} \left( \frac{d^C_d}{g_s}\right)_\Lambda
      + \frac{2}{9} \left( \frac{d^C_u}{g_s}\right)_\Lambda
      \,\right]\, ,
\end{equation}
where quantities  in the last bracket are  given by Eq.\ (\ref{CEDMf})
with the strong  coupling constant  $g_s$  and the $u$-  and $d$-quark
masses evaluated at the scale $\Lambda$.

In  Fig.\  2 we show  the dependence  of the EDMs  $d_e$ (solid line),
$d^C_n$ (dashed line), and  $d_n$  (dotted  line) on $\tan\beta$   and
$\mu$, for three different masses  of the would-be CP-odd Higgs  boson
$a$, $M_a = 100, 300, 500$ GeV. Since EDMs are mostly dominated by the
down-family fermions, i.e.,   the  electron and the down    quark, the
strong  $\tan\beta$   dependence is then  expected.    The EDMs depend
significantly on $\mu$ through the $a\tilde{f}^*\tilde{f}$ coupling in
Eq.\ (\ref{Lcp}).   Our analysis in  Fig.\ 2 clearly  shows that large
$\tan\beta$ scenarios, {\em i.e.}, $40< \tan\beta  < 60$ with $\mu,\ A
> 0.5$  TeV, $M_a \le  0.5$ TeV, and large  CP  phases are practically
ruled  out.  On  the other   hand, we find   that  for low $\tan\beta$
scenarios,    {\em e.g.}\  $\tan\beta \stackrel{\displaystyle <}{\sim}
20$,  the  two-loop Barr-Zee-type    contribution  to  EDMs   is  less
restrictive for natural values of parameters in the MSSM. Finally, the
results exhibit  an approximate linear  dependence on the  mass of the
$a$ boson  for the range of phenomenological  interest and validity of
perturbation theory, {\em i.e.}, for $0.1< M_a \stackrel{\displaystyle
  <}{\sim} 1$ TeV.

In conclusion, we have derived for the first  time {\em direct} limits
on the CP-violating parameters related to the third generation 
scalar-quarks, which originate from the experimental limits on the EDMs of
electron and neutron. 
These novel constraints  induced by the  SUSY version of the  two-loop
Barr-Zee graph will have an important impact on possible effects of CP
nonconservation   at collider  experiments,  on  dark-matter detection
rates,  and on studies of  the  baryonic asymmetry of  the Universe in
SUSY theories.

DC and AP wish to thank Fermilab Theory Group for hospitality.

\subsection*{ERRATUM}

There  is  a  normalization  factor  of  2  missing  in  the  analytic
expressions of  Eqs.\ (\ref{EDMf}) and (\ref{CEDMf}).   To be precise,
both  Eqs.\ (\ref{EDMf})  and (\ref{CEDMf})  must be  multiplied  by a
factor of 2.  As a result,  the numerical predictions for the EDMs are
by a factor 2 larger than those plotted in Fig.\ 2.  Finally, there is
a typographic  error in Eq.\  (\ref{xiq}): phase $\delta_b$  should be
replaced by $-\delta_b$.

\begin{figure}
   \leavevmode
 \begin{center}
   \epsfxsize=14.0cm
    \epsffile[0 0 539 652]{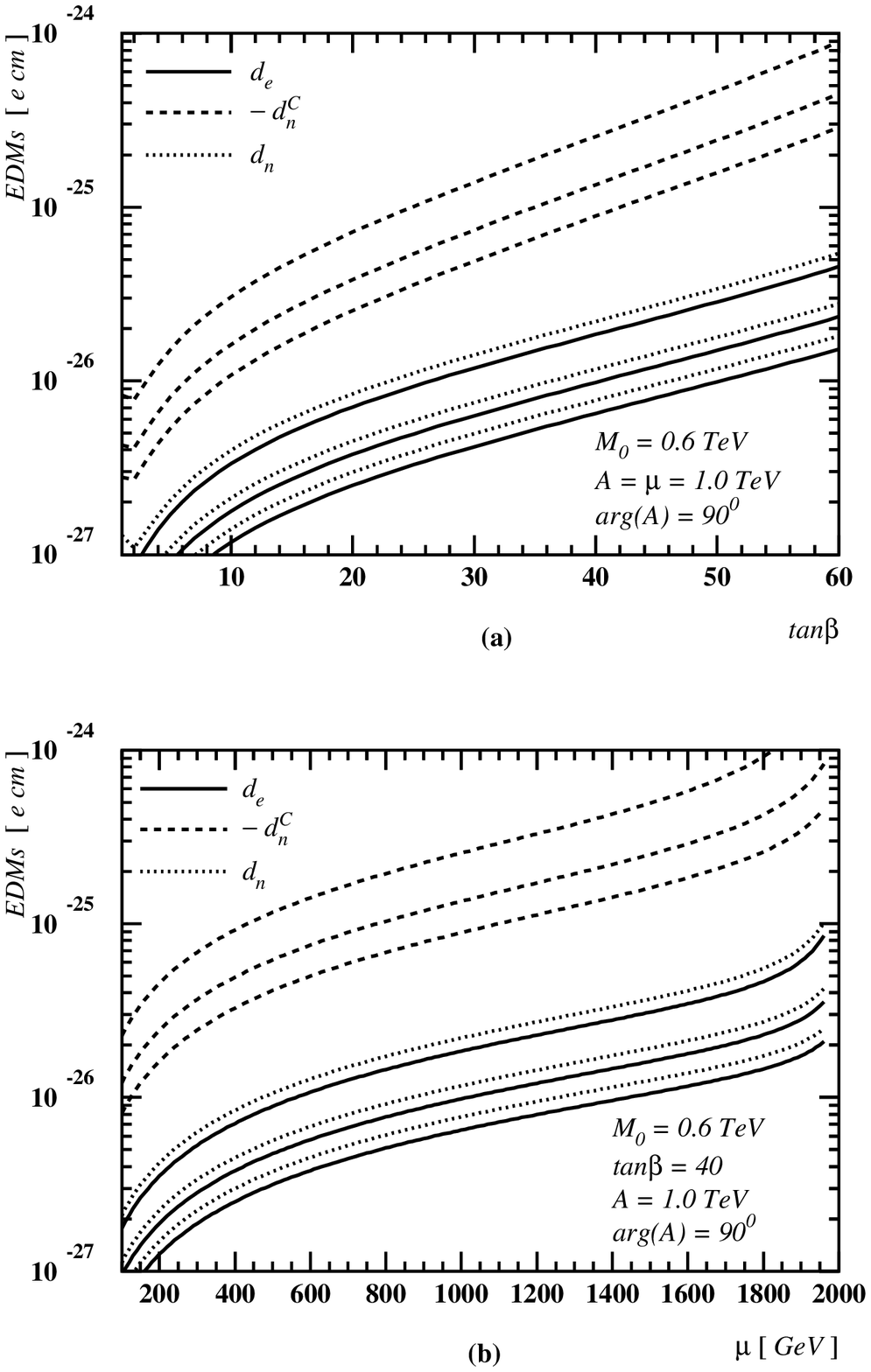}
 \end{center}
 \vspace{-0.5cm} 
\caption{Numerical estimates of EDMs. Lines of the same type
  from the upper to the lower one correspond to $M_a = 100,\ 300,\ 
  500$ GeV, respectively.}\label{f2}
\end{figure}

\end{document}